\documentclass[11pt]{article}
\begin{document}

\title{\LARGE \bf A twin paradox for `clever' students}
 \author{{{\bf S. K. Ghosal}\thanks{Author to whom all correspondences should be made.}, {\bf Saroj Nepal and  Debarchana Das}}\\
{{\em Department of Physics, North Bengal University}}\\
{{\em Dist. Darjeeling(WB), Pin: 734 013, India.}}\\
{{\em E-mail: ghosalsk@hotmail.com}}}

 \date{}
 
\maketitle

\thispagestyle{empty}
\vspace{2cm}
\begin{abstract}
\noindent
We discuss the twin paradox or the clock paradox under the small velocity approximation of special relativity. In this paper the traveller twin of the standard twin parable sets out with a {\em non-relativistic speed} for the trip leaving behind the 
stay-at-home one on earth and continues up to a distance and finally returns home with the same speed when the siblings can compare their ages or their initially synchronized wrist watches. The common knowledge that at small velocities the length contraction and time dilation effects of special relativity can be ignored so that the world becomes essentially classical, tends to lead to a paradox in connection with the twin problem, which a 
``clever'' student eventually discovers. After discussing and resolving the issue, some more related paradoxes have been presented. The resolution of all these problems provides some 
additional insight into the century old paradox.   \par  
\end{abstract}
\section {INTRODUCTION}
  The Lorentz transformation (LT) relates space-time coordinates of events between any two inertial frames in the relativistic world. The kinematical world of special relativity (SR) described by LT is endowed with the well known relativistic effects like length contraction and time dilation. The latter effect is of course responsible for the ``peculiar consequence'' that a clock $B$, initially synchronized with another stationary clock $A$ at one spatial point, after making a round-trip should fall out of step with the stationary one. Indeed $B$ would be seen to lag behind $A$ when they meet after the former's round-trip. The phrase ``peculiar consequence'' had originally been used by Einstein in 1905\cite{Einstein, Pesic} to describe the temporal offset effect, which subsequently was to be known as the clock paradox or the twin paradox\footnote{In the standard twin parable a traveller initially living on earth leaves behind his twin to take a trip to a distant star in a high speed rocket and returns in the same manner to his stay-at-home sibling to discover that they age differently. Depending on the context we will use the terms ``twin paradox'' or ``clock paradox'' interchangeably.}.\par
The peculiar or counter-intuitive consequence becomes paradoxical when one argues that since kinematically $A$ is also making a round-trip from the perspective of $B$ and LT predicts reciprocal time dilation of moving clocks, it is the clock $A$ which should run slow compared to the clock $B$ and the former would be seen to lag behind the latter when they meet, while the temporal offset between two clocks (or the age difference between twins) when compared at one spatial point after their reunion should be independent of the perspectives.\par
At one point in the history of science, the term ``paradox'' for this counter-intuitive problem was as if an euphemism for ``contradiction'', which was being used against the relativity theory by the opponents\cite{Builder1}. In these hundred years, students of relativity theory have become more matured and they now know that there is no paradox as such and the contradiction is only apparent. There is a basic asymmetry between the states of motion of the clocks---one is inertial and the other undergoes direction reversing acceleration and is therefore non-inertial and hence the asymmetric outcome in the difference of readings of the two clocks (asymmetric aging of the twins) does not in any way violate any matter of principle. They also know that although for the paradox to be well-posed, one of the clocks at least must reverse its direction during its journey (in order to facilitate comparison of clocks at one spatial point), the acceleration {\it per se} cannot play any role in calculating the temporal offset between the clocks from the {\em perspective of the inertial frame $A$}, since the duration of the turn around process of $B$ can be made arbitrarily small compared to that for the rest of its trip. (Some of them might still have doubt in this regard since the assumption of small turn-around time also implies a large acceleration of $B$, but this doubt will soon be removed once they read the articles by Gruber and Price (1997)\cite{Gruber} and Boughn(1989) \cite{Boughn}.) The understanding of the fundamental asymmetry between the frames of the clocks $A$ and $B$ removes the essential paradoxical element of the problem qualitatively (as to why the particular clock $B$ is seen to run slow); however in order to completely resolve the issue one should also be able to demonstrate that $B$ predicts the same time-offset between the clocks, that $A$ calculates using the simple time dilation effect of SR. Here we will assume that a ``clever'' student also knows how to do it. For these students of course the ``twin paradox'' does not exist.\par
We here devise a clock paradox for these learned students in the spirit of a conjurer who often reserves a sleight of hands item in his repertoire for the ``clever'' spectators who eventually get baffled by the outcome of the feat only as a result of their assumed ``knowledge'' of the secrets.\par
In the next section (Sec.2) we will pose the paradox and also provide the resolution. There will be some interesting ramifications of the paradox which will be discussed in section 3. The paradox with all its ramifications will hopefully provide a lot of insight into the hundred years old clock paradox along with some related special relativistic issues.

\section{THE PARADOX AND ITS RESOLUTION}
When relative velocities $v$ are very small compared to the speed of light $c$, such that ${v^2}/{c^2}$ terms can be neglected in comparison to unity, the so called Lorentz factor $\gamma= (1-{v^2}/{c^2})^{-1/2}$ can be assumed to be equal to $1$ and hence there is no contraction of rods or time dilation of moving clocks in this approximation. This is in conformity with the classical kinematics. These two typically relativistic effects do not show up unless $v$ is large. For small velocities therefore, the world is expected to be essentially classical or non-relativistic. However contrary to the common belief\footnote{The belief is often typically expressed as ``since classical physics does work in everyday life, it is essential that, for small $v$, the Lorentz transformation collapses to Galilean transformation'' \cite{Kacser}.}, for such small velocities, LT does not go over to the Galilean transformation (GT), instead it becomes the so called approximate Lorentz transformation (ALT) \cite{Landau,Zeitschrift,Mauro}:
\begin{equation}
\begin{array}{l}
x=x_0 -vt_0,\\
t=t_0-{vx_0}/{c^2},
\end{array}\\
\end{equation}
and for its inverse
\begin{equation}
\begin{array}{l}
x_0=x+vt,\\
t_0=t+{vx}/{c^2}.
\end{array}\\
\end{equation}
The transformations (1) and (2) are obtained by putting $\gamma=1$ in LT or in its inverse. The coordinates $(x_0,t_0)$ and $(x,t)$ refer to the space-time coordinate of the reference frames $S_0$ and $S$ respectively, where $S$ is assumed to move with uniform velocity $v$ along the common positive $x-$direction.\par
One should not feel disturbed by the presence of the phase (space dependent) terms in the time transformations of Eqs. (1) and (2). Indeed these terms cannot be dropped in the approximation since, for any preassigned small velocity $v$, the distance $x_0(x)$ from the origin of $S_0(S)$ can be taken to be arbitrarily large and hence the terms $v{x_0}/{c^2}$ in Eq. (1) or $v{x}/{c^2}$ in Eq. (2) may not be neglected. Because of the presence of the space dependent term in the time transformation, like in SR any two simultaneous events separated by a distance $\Delta x_0$ in $S_0$ will be non-simultaneous by the amount ${v\Delta x_0}/{c^2}$, as observed from $S$. The similar lack of agreement in simultaneity of spatially distant events holds in $S_0$ for the simultaneous events in $S$. Thus relativity of simultaneity, the well-known effect of SR, is preserved in the small velocity approximation. This happens, since the effect is linked to the convention of synchronization of clocks and it is unlikely that the synchrony character of SR will be altered by a mere small velocity approximation \cite{Zeitschrift}.\par
In relativity theory, distant clocks are assumed to be synchronized by light signals, stipulating as a convention, that the one-way-speed (OWS) of light is the same as the two-way-speed (TWS) of light. Note that TWS or the round-trip speed of light is a convention-free entity since for the measurement of TWS only one clock is employed and hence the problem of synchronization of spatially separated clocks does not enter. The stipulation of the equality of to and fro speeds of light (along any given direction) in any inertial frame is known as the Einstein synchrony. One can check by simple kinematical calculations that ALT still represents Einstein synchrony \cite{Zeitschrift}. The transformations (1) and (2) therefore represent a classical world (characterized by no time dilation and length contraction effects\footnote{The absence of these relativistic effects also directly follows from the ALT (Eqs.(1) and (2))}) with Einstein synchrony.\par
A not-so-clever student might now say that since there is no time-dilation of moving clocks with respect to both $S_0$ and $S$ according to ALT, the rocket-bound sibling (clock), after the round-trip, would not age less or more with respect to the stay-at-home one. The student might also assume that the prediction is independent of the perspectives of the observers and hence there is no twin paradox. The observation is also consistent with the common notion that mysteries, enigmas or peculiar consequences are the attributes of the relativity theory with its new philosophical (often counter-intuitive) imports and our mundane classical world is devoid of such things.\par
However, here an informed student may fall into a difficulty. We assume that this informed student knows the fact that the twin paradox arises ``solely out of the elementary mistake of utilizing, in a single calculation, quantities expressed in the measures of two different inertial reference systems\cite{Builder2}.''  Suppose $S_0$, with coordinates ($x_0, t_0$) represents the inertial frame of the stationary clock $A$. The rocket-bound clock $B$ is a non-inertial frame $K$ which may be assumed to be composed of two different inertial frames in the abrupt turn-around scenario. Let us call these frames $S$ with coordinates ($x, t$) for the onward journey and $S'$ with coordinates ($x', t'$) for the return journey of $B$. The coordinate transformation between $S_0$ and $S'$ is to be written as,
\begin{equation}
\begin{array}{l}
x'=x_0 +vt_0,\\
t'=t_0+{vx_0}/{c^2},
\end{array}\\
\end{equation}
and for the inverse,
\begin{equation}
\begin{array}{l}
x_0=x'-vt',\\
t_0=t'-{vx'}/{c^2},
\end{array}\\
\end{equation}
where we have assumed as usual, that the speed of $B$ for its forward and reverse journeys remain the same.\par
According to ALT the observer $A$ predicts no differential aging i.e the clock $B$ will agree with that of $A$ after the former's round-trip. Let us assume for future use that this round-trip time is $2T_0$. We shall now see however that according to the same transformation, $B$ will predict a different result, although with respect to both $S$ and $S'$, the $A$-clock (stationary with respect to $S_0$) does not suffer any time dilation.  Consider the coordinates ($X_0, T_0$) correspond to the event when the turn-around of the $B$-clock takes place. Assume that the corresponding coordinates in $S$ and $S'$ are given by ($X, T$) and ($X', T'$) respectively. Now $T$ and $T'$ are related to $X_0$ and $T_0$ by the time transformations of Eqs. (1) and (3) respectively:
 \begin{equation}
T=T_0 -{vX_0}/{c^2},
\end{equation}
\begin{equation}
T'=T_0+{vX_0}/{c^2},
\end{equation}
where $X_0$ now refers to the length $L$ of $B$'s journey for its outward trip. The difference of these times, $T'- T= \delta t_{gap}$ is given by,
\begin{equation}
\delta t_{gap}= {2vL}/{c^2},
\end{equation}
where the subscript ``gap'' refers to the synchronization gap at the event between the clock readings of the two inertial frames where the Einstein synchrony has been employed to synchronize their own coordinate clocks. With respect to $S$ (or $S'$) a clock at rest in $S_0$ does not suffer the usual time dilation, which is evident from the time transformations Eqs. (1) or (3) which, in the differential form read,
\begin{equation}
\Delta t= \Delta t_0-{v\Delta x_0}/{c^2},
\end{equation}
\begin{equation}
\Delta t'= \Delta t_0+ {v\Delta x_0}/{c^2}.
\end{equation}
Clearly the ``no time dilation'' result follows from the above equations {\it separately} when one puts the condition that $\Delta x_0$ is zero for the clock under consideration. Indeed this leads to the cavalier conclusion that the time taken by the stationary observer for its round-trip as interpreted by the rocket-bound observer ($B$), $\Delta t_{A}(B)$\footnote{Here and for the rest of the text we follow, for convenience, a notation scheme where $\Delta t_{X}(Y)$ stands for the round-trip time of $A$ or $B$ (as the case may be) that may have been recorded in the $X$-clock according to the perception or interpretation (write or wrong) of the observer $Y$.} should be the same as the $B$-clock time $\Delta t_{B}(A)$ as interpreted from $A$ for the former's round-trip. However the careful student understands that this sort of argument does not take care of the fact that the two separate inertial frames $S$ and $S'$ cannot be meshed to form the non-inertial frame $K$ of $B$, unless one takes into account the synchronization gap given by Eq. (7). Hence the correct round-trip time of $A$ as seen from $B$, $\Delta t_{A}^{true} (B)$, can be obtained by adding $\delta t_{gap}$ with the $B$-clock time (denoted by $\Delta t_{B}(B)$) in its own frame and then take care of time dilation (or its {\em absence}) of the $A$-clock with respect to $B$. Hence one may write in this case (where the time dilation factor is unity),
\begin{equation}
\Delta t_{A}^{true} (B) =\Delta t_{B}(B)+{2vL}/{c^2}.
\end{equation}
Note that the $B$-clock time, i.e the proper time of $B$, for the round-trip, according to the indicated notation is given by,
\begin{equation}
\Delta t_{B} (B)=\Delta t_{B}(A)= \Delta t_{A}(A)=2 T_0 .
\end{equation}
Where the first equality is evidently true since the proper time of $B$ does not differ with $A$'s interpretation of the $B$-clock time, as it is stationary in an inertial frame $S_0$, where there is no break in synchrony. The second equality holds since there is no time dilation and the last equality is the restatement of our assumption that the round-trip time recorded in $A$-clock is $2T_0$. Hence 
\begin{equation}
\Delta t_{A}^{true} (B) =2T_0+{2vL}/{c^2}.
\end{equation}  
 Therefore, $B$ predicts a temporal offset between the clocks
\begin{equation}
\delta t_{off} (B) =\Delta t_{A}^{true} (B)-\Delta t_{B} (B)={2vL}/{c^2}.
\end{equation}
This is {\em paradoxical}, since as we have explained, the observer $A$ does not predict any temporal offset between the clocks after their reunion. Interestingly this time difference (Eq. (13)), whose origin lies in the synchronization gap between two inertial frames, has nothing to do with the usual time dilation or the Lorentz factor. Indeed for any preassigned small value of $v$, the length of the round-trip, 2L can be chosen to have an arbitrarily large value so that the amount by which the two clocks fall out of step can always be made measurable. Thus a knowledgeable student now does not enjoy the bliss of a less informed one, who thinks that both $A$ and $B$ predict agreement of the clocks after their reunion.\par
The answer to this paradox, as a virtuoso might have already guessed, is not hard to seek. Observe that instead of emphasizing on the time-offset (measured by a clock) if one considers the age of $A$ as interpreted from $B$ and compares the same with his own age ($2T_0={2L}/{v}$) then from the relation (10) and (12)one obtains,
\begin{equation}
\Delta t_{A}^{true} (B)=({2L}/{v})(1+{v^2}/{c^2}).
\end{equation}
Hence it is observed that the extra aging as interpreted from $B$ is again a second order effect. For the standard twin paradox however, for small velocities the extra aging of $A$ can be neglected since, under the approximation one may not be able to discern the extra shade of grayness in the hair color of the stay-at-home twin in comparison to that of the traveller sibling. Indeed when we compare ages we do compare the absolute measures of time; in contrast for the comparison of ordinary clocks (standard 12 hour wrist watches say), instead of the absolute measure, we are often interested to know only how much one lags behind the other.\par
In the standard twin paradox scenario therefore the answer to the contradiction is that, the two observers {\em do agree} in their conclusions that there is no asymmetrical aging under the {\em stipulated approximation}. For the clock paradox however, the answer would be that, since the time-offset between the observers' clocks can always be a measurable (for large L), albeit a second order effect one cannot ignore the $\gamma$-terms in LT to start with; and once one includes them in the transformation, $A$ also predicts a non-zero temporal offset viz. $({2L}/{v})(1-\gamma^{-1})$, which comes out to be the same if calculated from $B$'s perspective taking the synchronization gap into account (to understand this one may follow the line of arguments leading to Eq. (13), but this time not ignoring the relativistic time dilation of $A$'s clock as seen from $B$). In other words one will have to follow the arguments leading to the resolution of the usual twin (clock) paradox.
\section{RAMIFICATIONS}
\subsection{Ramification I}
The paradox does not end here. From the foregoing discussion it appears that to avoid any contradiction between the two observers conclusions regarding the time-offset, one must use the full LT. One may now ask, in view of the contradiction encountered in using ALT (forgetting its history), ``Is there anything wrong with the transformation equation (1) itself? Can it not represent even a hypothetical kinematical world with the presence or the absence of length contraction and time dilation and a given (standard) synchronization scheme?'' In fact if one assumes that ALT (Eq. (1)) by itself (not as an approximation of LT) represents some kinematical world, the contradiction cannot be answered in terms of any error in ignoring the second order (in $v/c$) effect while posing the paradox, since in this case one is not dealing with any approximations at all to start with. The answer to this question is even more interesting. Indeed the transformation (1) represents a kinematical world with the Einstein synchrony but Eq. (2) is not the inverse of Eq. (1), if one is interested in the second order effects and beyond. Indeed by inverting Eq. (1) one obtains,
  \begin{equation}
\begin{array}{l}
x_0=\gamma^2 (x+vt),\\
t_0=\gamma^2 (t+{vx}/{c^2}).
\end{array}\\
\end{equation}
Hence Eq. (1) and Eq. (15) represent the direct and inverse transformations between the coordinates of the inertial frames $S_0$ and $S$ respectively, which represent time dilation (with factor $\gamma^2$) but no length contraction with respect to $S_0$.\par
The pair of transformations (1) and (15), as we shall see below, {\em do not} lead to the contradiction (as Eqs.(1) and (2) did) that the two observers predict different time-offsets between their clocks after their reunion. We first note that the time dilation for the moving clock with respect to $S_0$ can be obtained from Eq. (15) as
\begin{equation}
\Delta t_{0}=\gamma^2 \Delta t,
\end{equation}
In order to arrive at this relation one puts in the differential form of Eq. (15) as usual, $\Delta x = 0$, for the clock stationary in $S$. Hence $B$-clock time can be obtained from the $A$-clock time by multiplying the latter by $\gamma^{-2}$. For the round-trip of $B$, the $A$-clock time is clearly given by,
\begin{equation}
\Delta t_{A} (A)={2L}/{v}.
\end{equation}
Hence the $B$-clock time is obtained as
\begin{equation}
\Delta t_{B} (A)=({2L}/{v})\gamma^{-2}.
\end{equation}
Therefore the time-offset from $A$'s perspective comes out to be
\begin{equation}
\delta t_{off}(A)=\Delta t_{A} (A)-\Delta t_{B} (A)={2Lv}/{c^2},
\end{equation}
which is the same as the offset obtained earlier from the perspective of $B$, $\delta t_{off}(B)$, solely in terms of synchronization gap between the frames $S$ and $S'$ at the event of the turn-around\footnote{Had the transformation been LT, one would further need to correct the result (Eq. 13) by taking also the effect of time dilation of the clock $A$ with respect to $B$.}. 
\subsection{Ramification II}
There is a further ramification to the issue. Suppose now instead of the transformation (1) one uses the form of Eq. (15) to represent the direct transformation of space-time coordinates of $S_0$ to that of $S$: 
\begin{equation}
\begin{array}{l}
x=\gamma^2( x_0 -vt_0),\\
t=\gamma^2( t_0-{vx_0}/{c^2}),
\end{array}\\
\end{equation}
the corresponding inverse 
\begin{equation}
\begin{array}{l}
x_0=x+vt,\\
t_0=t+{vx}/{c^2},
\end{array}\\
\end{equation}
represents the transformation of coordinates from $S$ to $S_0$. This is what happens if one makes the approximation $\gamma\approx1$ in the inverse LT (representing transformation of coordinates from $S$ to $S_0$) to obtain Eq. (21) first and then takes its inverse (Eq. (20)) algebraically (forgetting the approximation). From the point of view of $A$ which is again stationary in $S_0$, there is now no time dilation, hence from $A$'s point of view,  the clocks will not fall out of step after their reunion, i.e., one should have 
\begin{equation}
\delta t_{off} (A) =0.
\end{equation}
Here one may predict a contradiction since, according to the time transformation of Eq. (20), the observer $B$ perceives a time dilation of the $A$-clock. However, here also as we shall see a correct approach will provide, as before, an unequivocal prediction for the time-offset between the clocks. Note that in arriving at result (22) we have assumed that as seen from $A$, the round-trip time of $B$ as recorded by $A$-clock $\Delta t_{A}(A)=2L/v$ is the same as $\Delta t_{B}(A)$ which is nothing but the latter's proper time $\Delta t_{B}(B)$ (for contrast see Eqs. (17) and (18)). Let us write these relations for future reference
\begin{equation}
\Delta t_{A} (A)= 2L/v = \Delta t_{B}(A)= \Delta t_{B}(B).
\end{equation}
However $B$ cannot apply the time dilation formula to interpret the $A$-clock time from $\Delta t_{B}(B)$, directly, as there is a question of the ``synchronization gap'' between the two inertial frames $S$ and $S'$ which constitute the non-inertial frame $K$ of $B$. From $B$'s point of view, the synchronizing gap between $S$ and $S'$, following the previous arguments leading to Eq. (13) is now given by, 
\begin{equation}
\delta t_{gap} =({2vL}/{c^2})\gamma^2.
\end{equation}
Therefore before one calculates $A$'s time from $B$-clock's time $\Delta t_{B}(B)$ one must add, as before, $\delta t_{gap}$ to it. Let us denote this added value by $\Delta t_{B}^{+}(B)$, which using Eq. (23)is given by
\begin{equation}
\Delta t_{B}^{+} (B) =\Delta t_{B} (B)+\delta t_{gap}={2L}/{v}+({2vL}/{c^2})\gamma^2.
\end{equation} 
$A$-clock's time will be found by using the time dilation formula obtained from the time transformation (20) (by putting $\Delta x_0 = 0$ in its differential form, as usual):
\begin{equation}
\Delta t =\gamma^2\Delta t_{0},
\end{equation}
Note that the above formula is independent of the direction of $B$'s journey i.e whether $B$ is in $S$ or in $S'$. Hence using the formula (26),
\begin{equation}
\Delta t_{A}^{true} (B) =\gamma^{-2}\Delta t_{B}^{+}(B)=\gamma^{-2}[{2L}/{v}+({2vL}/{c^2})\gamma^2],
\end{equation}
which after simplification gives,
\begin{equation}
\Delta t_{A}^{true} (B) ={2L}/{v}.
\end{equation}
Observe that the right hand side of Eq. (28) is nothing but $\Delta t_{B}(B)$ (see Eq. (23)). Hence the time-offset from the perspective of $B$ given by,
\begin{equation}
\delta t_{off} (B) =\Delta t_{A}^{true}(B)-\Delta t_{B} (B),
\end{equation}
also turns out to be zero, thus dissolving the contradiction.
\section{CONCLUDING REMARKS}
Finally one may wonder that the pairs of transformation Eqs. (1) and (15) and Eqs. (20) and (21) are essentially the same transformations, where only in the latter pair $v$ has been replaced by $-v$, yet the two pairs yield different results for the differential aging of the twins. The second pair predicts no temporal offset while for the former there is a differential aging.\par
The issue is more subtle and is intimately connected with the question as to why the ``broad principle of relativity'' of motion fails leading to asymmetric aging \cite{Muller} of the twins. In the last two cases the role of $S_0$ and $S$ has been interchanged. For $B$-clock's forward journey the inertial frame of reference attached to the clock is $S$, but for the return journey it is $S'$. The break in (Einstein) synchrony occurs in $K$ i.e between the inertial frames $S$ and $S'$ and not in $S_0$. Although with respect to $B$, the observer $A$ stationary in $S_0$ executes a to and fro motion, it is implicitly assumed that there is {\em no} break in synchrony in $S_0$. We have thus given, $S_0$ a special status. Since this is an inertial frame, the two way speed of light (a synchrony independent quantity) is isotropic in it and besides the round-trip time for a given circuital path in this frame is a constant in time\footnote{Indeed in an accelerated frame the result is different. For example, the effect of acceleration on the round-trip time of a light signal is evidenced in the ordinary \cite{Post} or linear Sagnac effects \cite{Ehrenfest,Wang} and the effect can be calculated using the simple kinematics of SR \cite{Post,Ehrenfest,Dieks,Selleri}.}. In this frame the light signal synchronization using Einstein's convention is an unambiguous procedure. But in $K$ i.e in the frame attached to $B$, which executes a to and fro motion with respect to $S_0$, the synchronization can be done consistently only locally i.e in the inertial frames $S$ and $S'$ separately, however, it can be shown that it is kinematically impossible to follow Einstein's procedure for synchronization globally in $K$ \cite{Anisotropy}\footnote{For example one finds a discontinuity problem in following Einstein Synchronization along the circumference of a rotating disk, which is often regarded as the real cause of the Sagnac effect \cite{Rizzi}. The synchronization gap or the desynchronization issue in a rotating frame has also been discussed in details in a recent paper by Cranor et-al in the context of their circular twin paradox \cite{Cranor}. }. Thus having understood the preferred status of $S_0$ in this sense, we can distinguish the two worlds represented by the pairs of transformation equations (1) and (15) and, (20) and (21) in terms of kinematical effects like length contraction and time dilation with respect to $S_0$ only. Therefore, since the role of $S_0$ and $S$ have been interchanged in the last two pairs of transformation equations, it is no longer surprising that the worlds should differ qualitatively by predicting different differential agings.\par
The above discussions make it clear once more that at the heart of the twin paradox in relativity lies the clock synchronization issue or in C. G. Darwin's \cite{Darwin} words ``the mysteries of time-in-other-places''. This is even true as we have seen, in the 
two aforementioned kinematically different (non-classical as well as non-relativistic) hypothetical worlds, where also one has employed Einstein's synchrony in the reference frames.\par
There are many elegant treatments of the paradox in different pages of various journals but the question as to what mistake in the reasoning leads to the paradox in the first place often remains obscured \cite{Muller}. The present treatment of the paradox 
(although of different variants) underpins the fact that the mistake in the reasoning in posing the counter-intuitive problem is purely kinematical and hence the error is to be corrected only kinematically. Surprisingly generations of students have been mislead to believe that the solution of the twin paradox lies in the realm of general relativity (GR)\footnote{Even Einstein believed (as a second thought) that ``the resolution of the paradox lay outside the realm of special relativity''\cite{Muller}. For some historical accounts and comments on the issue see a recent article by Peter Pesic \cite{Pesic}.}.  However introduction of GR for solving the issue in an essentially flat space-time (with vanishing Riemann tensor) is ``decidedly misleading'' \cite{Redhead}. This is now more apparent in the context of the questions addressed here since problems arising out 
of fallacious arguments should be cured only by carefully correcting the faulty reasonings and not by invoking any new physical effects\footnote{A recent article \cite{Twinp-ep} in the context of the use of equivalence principle of GR to solve the twin paradox \cite{Harpaz}, has shown (in essence) that such an exercise is equivalent to resolving the problem by using relativistic kinematics only.}.

{\bf Acknowledgment:}
One of the authors (SN) would like to thank the Tribhuvan University and the University Grants Commission, Nepal for providing a research fellowship at NBU.

 \end{document}